\documentclass[twocolumn,english,aps,superscriptaddress]{revtex4}

\usepackage{amssymb}
\usepackage{amsmath}
\usepackage{colordvi}
\usepackage[colorlinks]{hyperref}
\usepackage{amsthm}
\usepackage{subeqnarray}
\usepackage{cases}
\usepackage{enumerate}
\usepackage{graphicx}
\usepackage{epsfig}
\usepackage{epstopdf}
\usepackage{subfigure}
\usepackage{color}
\usepackage{url}
\usepackage{verbatim}
\usepackage{mathrsfs}

\usepackage{braket}
\renewcommand{\vec}[1]{\mathbf{#1}}

\begin{document}

\title{Plasma-enhanced interaction and optical nonlinearities of Cu$_2$O Rydberg excitons}
\author{Valentin Walther}
\email{valentin.walther@cfa.harvard.edu}
\affiliation{Department of Physics and Astronomy, Aarhus University, Ny Munkegade 120, 8000 Aarhus C, Denmark}
\affiliation{ITAMP, Harvard-Smithsonian Center for Astrophysics, Cambridge, Massachusetts 02138, USA}
\affiliation{Department of Physics, Harvard University, Cambridge, Massachusetts 02138, USA}
\author{Thomas Pohl}
\affiliation{Department of Physics and Astronomy, Aarhus University, Ny Munkegade 120, 8000 Aarhus C, Denmark}

\begin{abstract}
 We theoretically investigate the nonlinear optical transmission through a cuprous oxide crystal for wavelengths that cover the series of highly excited excitons, observed in recent experiments. Since such Rydberg excitons have strong van der Waals interactions, they can dynamically break the conditions for resonant exciton creation and dramatically modify the refractive index of the material in a nonlinear manner. We explore this mechanism theoretically and determine its effects on the optical properties of a semiconductor for the case of degenerate pair-state asymptotes of Rydberg excitons in Cu$_2$O. Upon analyzing the additional effects of a dilute residual electron-hole plasma, we find quantitative agreement with previous transmission measurements, which provides strong indications for the enhancement of Rydberg-induced nonlinearities by surrounding free charges. 
\end{abstract}

\maketitle

\begin{figure}[t]
\begin{center}
 \includegraphics[height=.20\textwidth]{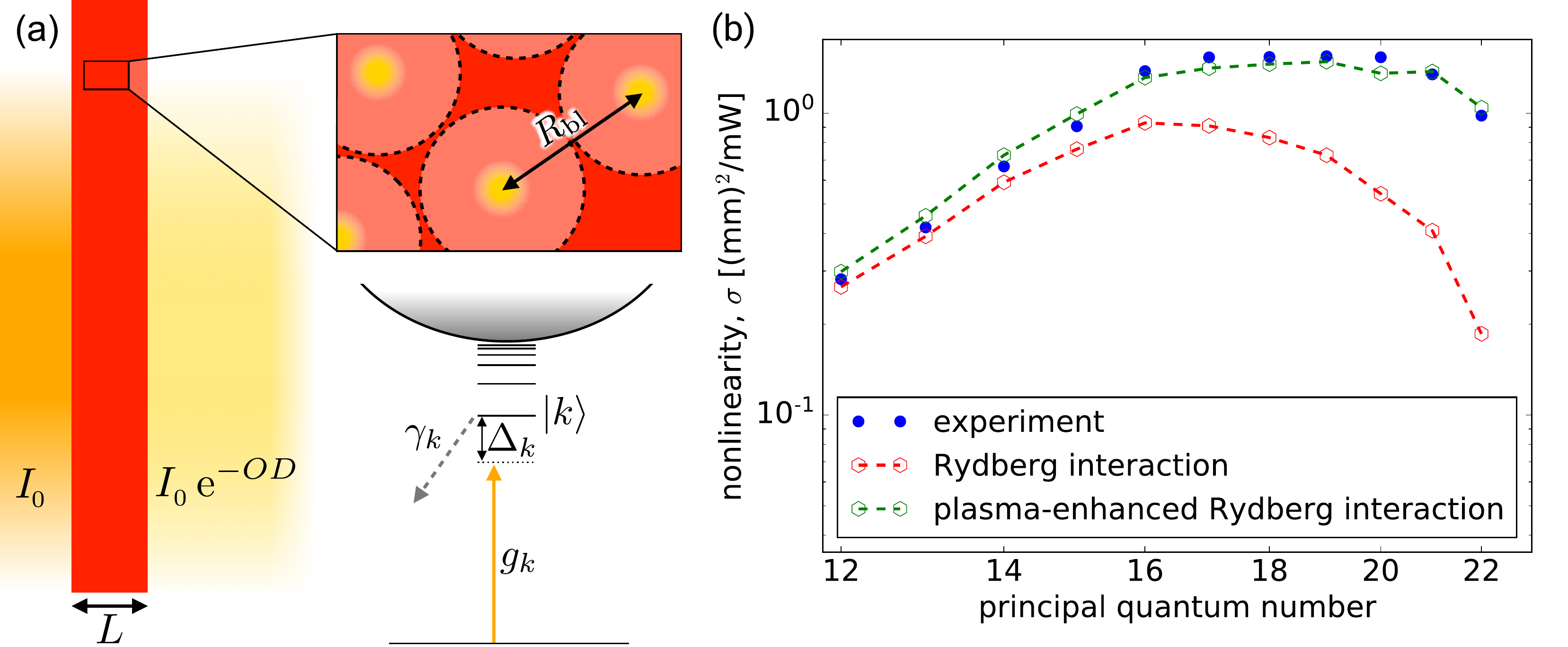}
\end{center}
\caption{The transmission of a laser field with incident intensity $I_0$ through a Cu$_2$O crystal of length $L$ can be strongly affected by the resonant optical response of exciton states $|k\rangle$. Hereby, the excitons are generated with light-matter couplings $g_k$, and their spectral properties are determined by the corresponding frequency detunings $\Delta_k$ and decay rates $\gamma_k$. A large optical nonlinearity arises from the strong van der Waals interaction between excitons in highly excited Rydberg states, which inhibits the simultaneous creation of excitons with a blockade radius $R_{\rm bl}$ and thereby profoundly alters their absorptive response to the incident light. b) The calculated optical nonlinearities based on this mechanism are in quantitative agreement with existing measurements \cite{kazimierczuk2014giant}, upon including the effects of plasma screening on high lying Rydberg exciton states.}
\label{fig1}
\end{figure}

Excitons, bound quasi-particle states of an electron and a hole, dominate the optical response of many semiconductors below the bandgap. Since their first discovery \cite{gross1952, gross1956} almost 70 years ago, excitons in Cu$_2$O have continued to attract substantial interest, including recent studies of their behavior in external fields \cite{schweiner2017, rommel2018, heckoetter2017, zieliska2019}, their interactions with phonons \cite{schoene2017}, their response to strain traps \cite{krueger2018} and their peculiar optical selection rules \cite{thewes2015, konzelmann2019}. 
This renewed attention to Cu$_2$O excitons originates from the spectacular discovery of excitonic Rydberg states, which could be resolved with unprecedented precision and up to extreme principal quantum numbers of $n=25$ \cite{kazimierczuk2014giant}. One of the experimental surprises \cite{kazimierczuk2014giant} has been a declining absorption with increasing laser intensity that occurs already for remarkably weak light fields. The onset of this effect shifts to lower and lower intensity with increasing principal quantum number of the Rydberg state, hinting at an underlying interaction mechanism of the formed excitons. Indeed, Rydberg excitons feature strong van der Waals interactions \cite{walther2018interactions} that rapidly increase with $n$, and can therefore give rise to an interaction-induced exciton blockade that is analogous to the Rydberg blockade observed in cold atomic systems \cite{gaetan2009,urban2009,schauss2012,busche2017}.

Here, we present a quantitative theory of the measured Rydberg-exciton spectra in Cu$_2$O and show that the observed line suppression can be understood in terms of a sizable optical nonlinearity that arises from the strong van der Waals interaction between the excitons. In particular, we demonstrate that screening by a residual electron-hole plasma in the semiconductor \cite{heckoetter2018} tends to strengthen the excitonic Rydberg-state interactions and thereby further enhances the optical nonlinearity. The calculations based on such plasma-enhanced interactions predict an exciton blockade at mesoscopic scales, which quantitatively describes existing absorption measurements \cite{kazimierczuk2014giant} over the full range of observed Rydberg-state levels (see Fig.\ref{fig1}). The understanding of blockade effects and demonstration of enhanced exciton interactions in semiconductors provides an essential step towards their use for nonlinear and quantum optics \cite{walther2018giant, khazali2017} and explorations of many-body phenomena \cite{poddubny2019topological}. 
 
\begin{figure}
\begin{center}
 \includegraphics[height=.37\textwidth]{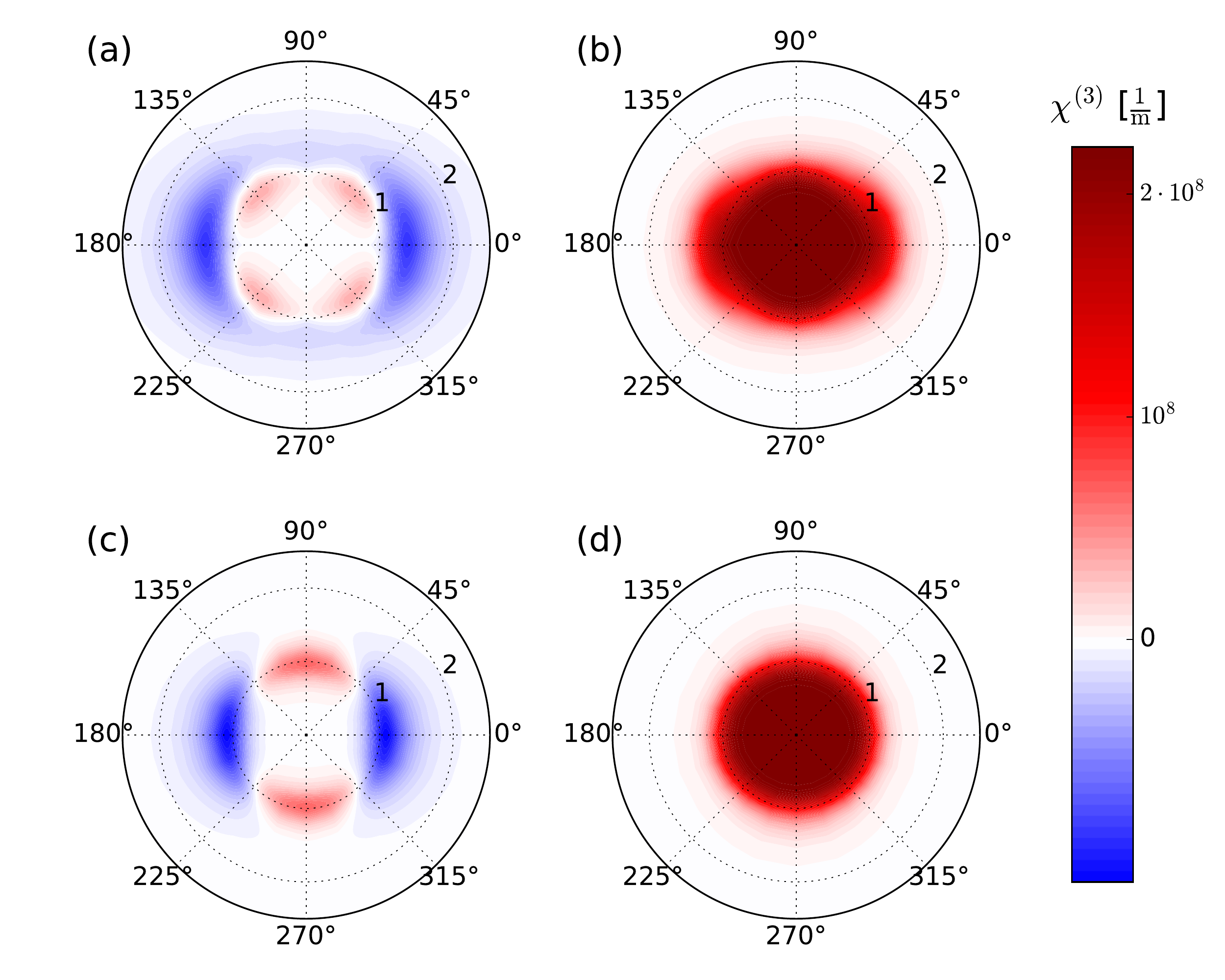}
\end{center}
\caption{Nonlinear and nonlocal optical susceptibilities $\chi^{(3)}$ for resonant excitation of the $15p-15p$-asymptote as a function of the exciton-exciton distance (radial coordinate in $\mu$m) and angle $\theta$ between the distance vector and the polarization axis of the applied light laser field for linear (a-b) and circular polarisation (c-d). Nonlinear refraction is represented by the real part $\Re(\chi^{(3)})$ (a,c), while the nonlinear absorption coefficient, $\Im(\chi^{(3)})$, is shown in panels (b,d). Symmetry under exciton exchange leads to the depicted fourfold mirror symmetry.}
\label{fig2}
\end{figure}

We begin by considering the transmission of a weak coherent light field $\mathcal{E}(\mathbf{r})$ through a Cu$_2$O crystal of thickness $L$ with approximately flat and parallel surfaces [see Fig.\ref{fig1}(a)]. The slowly-varying field amplitude $\mathcal{E}(r)$ is defined such that it relates to the intensity via $I(\mathbf{r}) = \hbar \omega c/\bar{n} |\mathcal{E}(\mathbf{r})|^2$ with a photon energy $\hbar \omega$ and a phase velocity $c/\bar{n}$ that is reduced from the free-space value $c$ by the dielectric constant $\bar{n}=\sqrt{\epsilon}=2.74$. Light propagation through the material can then be described by
\begin{align} \label{eq:field}
 \partial_{t} \mathcal{E}(\vec{r}) + \frac{c}{\bar{n}} \partial_{z}\mathcal{E}(\vec{r}) &= - i \sum_{k}\frac{g_{k}}{\bar{n}^2}\hat{X}_{k}(\vec{r}),
\end{align}
where the left-hand side captures the free evolution of the field envelope, while the term on the right-hand side accounts for the optical generation of excitons with a light-matter coupling strength $g_k$. The bosonic operators $\hat{X}_k^\dagger$ create these excitons in a given state with a set of quantum numbers that are labelled by the index $k$. Since typical transverse beam profiles are broad, diffraction is of minor importance and can be neglected in Eq. (\ref{eq:field}). The exciton energy $\hbar \omega_k$ and the laser frequency $\omega$ define the detuning $\Delta_k=\omega-\omega_k$ from a given exciton state whose complex width $\Gamma_k = \gamma_k-2i\Delta_k$ (Fig.\ref{fig1}a) also contains the decoherence rate $\gamma_k$. The decoherence rate typically describes the combined result of spontaneous decay, phononic coupling \cite{toyozawa1959, jolk1998} and inhomogenous sources of broadening. 

Excitons of the yellow series in Cu$_2$O are formed from the highest valence band and the lowest conduction band, which have the same parity. Therefore, excitonic $s$-states are dark ($g_{s} = 0$) and $p$-states dominate the exciton series \cite{kazimierczuk2014giant}. While for low principal quantum numbers and correspondingly small excitonic states the cubic lattice structure can admix some coupling strength to the $f$- and $h$- excitons \cite{thewes2015}, the selection rules for the highly-excited states, considered here, ensure a virtually exclusive excitation of Rydberg $p$-states. By controlling the polarization of the incident laser light one can thus selectively excite different angular momentum states within a given $p$-state manifold, which, in turn, affects the resulting nonlinearities, as we will show below.

The strong interaction between Rydberg excitons at long distances arises predominantly from transition dipoles that couple a given pair state $|jk\rangle$ of two excitons at positions $\vec{r}$ and $\vec{r}^\prime$ to other pair states $|j^\prime k^\prime\rangle$ at the same positions with matrix elements $V_{jk,j^\prime k^\prime}(\vec{r}-\vec{r}^\prime)$. The resulting correlated dynamics of multiple excitons, driven by the incident light and their mutual interactions, can therefore be described by
\begin{equation}\label{eq:ex_nonlinear}
\begin{aligned} 
& \partial_{t} \hat{X}_k(\vec{r}) = -\frac{\Gamma_{k}}{2}\hat{X}_{k}(\vec{r}) - i g_k\mathcal{E}(\vec{r}) \\
 &- i \int d \vec{r}^{\prime} \sum_{j,j^\prime,k^\prime} V_{j k,j^\prime k^\prime}(\vec{r}-\vec{r}^{\prime})\hat{X}_{j}^{\dagger}(\vec{r}^{\prime})\hat{X}_{j^\prime}(\vec{r}^{\prime})\hat{X}_{k^\prime}(\vec{r}),
\end{aligned}
\end{equation}
which together with Eq.~(\ref{eq:field}) determines the coupled dynamics of light and excitons and can be used to calculate the transmission and absorption properties of the semiconductor. 

In the linear regime, the optical absorption length, $\ell_{\rm abs}$, is readily obtained from the steady state of Eqs.~(\ref{eq:field}) and ~(\ref{eq:ex_nonlinear}) in the absence of interactions ($V_{j k,j^\prime k^\prime}=0$). This readily yields the absorption length $\ell_{\rm abs}^{-1}=-\frac{4}{\bar{n}c}\sum_k\frac{g_k^2\gamma_k}{|\Gamma_k|^2}$ and corresponding optical density $OD=L/\ell_{\rm abs}$ of the material at each absorption peak, corresponding to a resonance with a given exciton state $|k\rangle$. Analyzing the interacting case turns out far more involved as it requires the inclusion of strong dipole interactions and exciton correlations that are not amendable to simplifying perturbative approaches.

Following the general strategy outlined in \cite{sevincli2011,walther2018giant}, we start from the set of evolution equations~(\ref{eq:field}) and ~(\ref{eq:ex_nonlinear}) which are coupled to the correlator $\hat{X}_{j}^{\dagger}(\vec{r}^{\prime})\hat{X}_{j^\prime}(\vec{r}^{\prime})\hat{X}_{k^\prime}(\vec{r})$. Its evolution equation, obtained by applying the chain rule to Eq.~(\ref{eq:ex_nonlinear}), contains single-exciton terms $\hat{X}_{i^\prime}^{\dagger}(\vec{r}^{\prime})\hat{X}_{i}(\vec{r}^{\prime})$, and two-exciton terms $\hat{X}_{i}(\vec{r}^{\prime})\hat{X}_{j}(\vec{r})$ and $\hat{X}_{i^\prime}^{\dagger}(\vec{r}^{\prime})\hat{X}_{j}(\vec{r})$, as well as higher order operator products that describe correlations between three excitons. This procedure yields an infinite hierarchy of successively coupled operator equations, which needs to be truncated in order to obtain a tractable computation scheme. However, it turns out that the leading order nonlinearity, which is proportional to the square of the field amplitude $\mathcal{E}$, can only stem from excitonic two-body terms, while correlators of three or more excitons exclusively contribute to higher-order nonlinearities. Therefore, we can neglect three-body and higher-oder terms, and still obtain an exact solution for the third-order nonlinearity that accounts for arbitrarily strong exciton interactions, $V_{j k,j^\prime k^\prime}$. 

To facilitate such a calculation, we first transform the resulting set of two-body equations into a coordinate frame in which the exciton-exciton distance vector $\vec{r}-\vec{r}^\prime$ defines the quantization axis for the excitonic states $|k\rangle$.
Such a transformation is accomplished via Wigner-$d$ matrices $d_{i,k} (\theta)$ \cite{wigner1959} with an angle $\theta$ between $\vec{r}-\vec{r}^\prime$ and the original quantization axis defined by the incident laser field in the laboratory frame (see App.~\ref{app:eq_motions}). In this "molecular frame", the dipole-dipole coupling in Eq.~(\ref{eq:ex_nonlinear}) preserves the total angular momentum of the exciton pair-states, such that we can directly use the van der Waals interaction potentials $U_{\mu}$ and corresponding bi-exciton states $|\mu\rangle$ \cite{walther2018interactions} obtained by diagonalizing the underlying dipole-dipole interaction Hamiltonian. Transforming our operator equations into such a basis in which the interactions are diagonal, then permits to determine the steady state of $\langle\hat{X}_k\rangle$ and thereby yields the nonlinear field-propagation equation
\begin{equation}\label{eq:field_nonlinear}
\begin{aligned} 
i\partial_z \mathcal{E} (\vec{r}) = \chi^{(1)} \mathcal{E} (\vec{r}) + \int d\vec{r}^\prime \chi^{(3)}(\vec{r}-\vec{r}^\prime) |\mathcal{E}(\vec{r}^\prime)|^2 \mathcal{E}(\vec{r})
\end{aligned}
\end{equation}
with explicit expressions (see App.~\ref{app:eq_motions}) for the optical suceptibilities $\chi^{(1)}$ and $\chi^{(3)}$ in terms of the laser parameters, the interaction potentials $U_\mu$ and compositions of the corresponding bi-exciton states $|\mu\rangle$. This equation captures the nonlinear light propagation to leading order in the field intensity and for arbitrarily strong exciton-exciton interactions.

Fig.\ref{fig2} shows typical examples for the real and imaginary part of the obtained nonlinear kernel $\chi^{(3)}$, which can be understood as an effective photon interaction and a nonlinear absorption coefficient, respectively. The depicted surfaces reveal a significant influence of the laser polarization and a pronounced angular structure. Both properties stem from the combined contribution of several distinct non-spherical exciton-interaction potentials \cite{walther2018interactions} with degenerate asymptotes, corresponding to different angular momentum states of the interacting exciton pairs.

\begin{figure}[t]
\begin{center}
 \includegraphics[height=.35\textwidth]{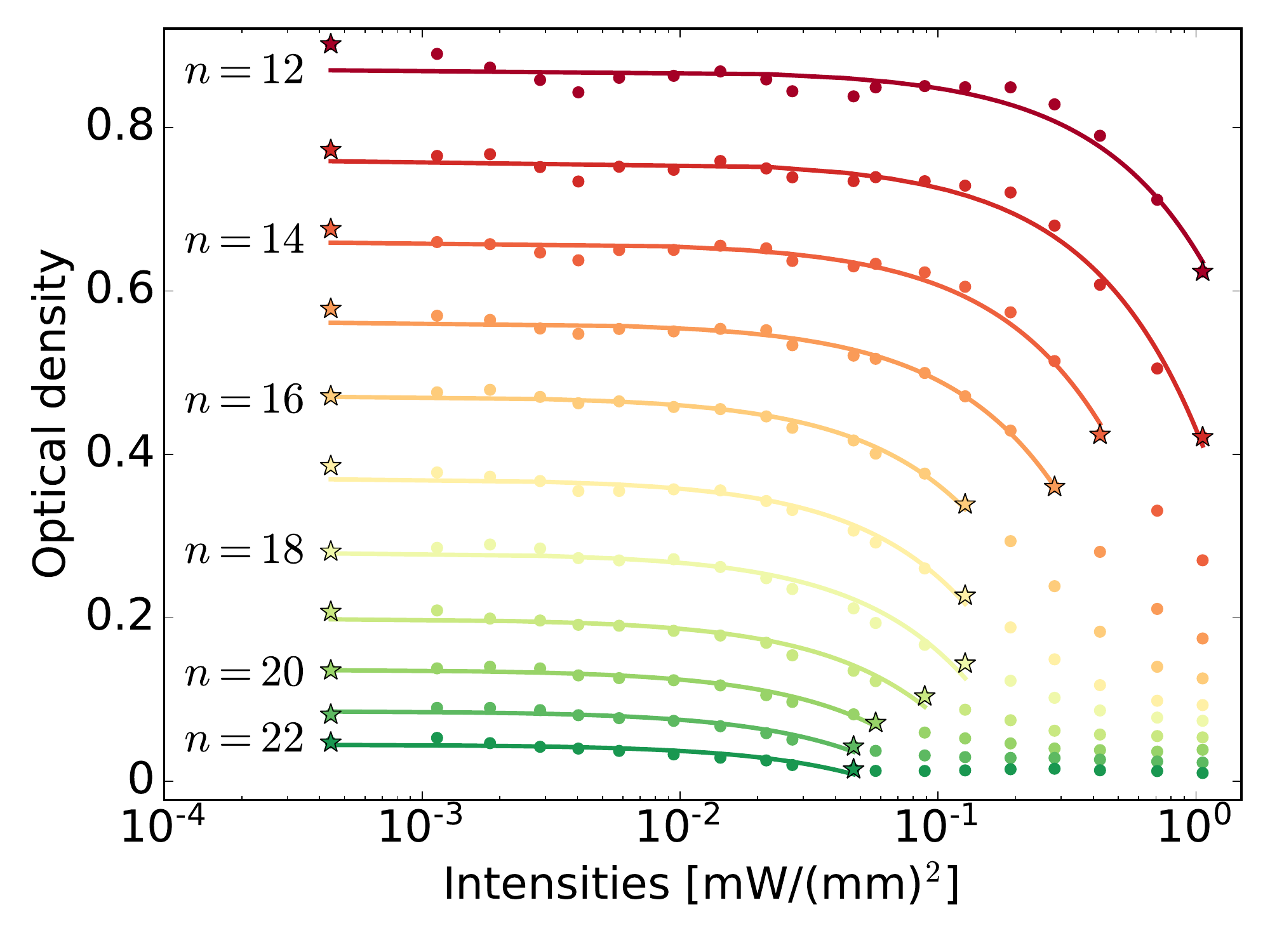}
\end{center}
\caption{Experimentally measured \cite{kazimierczuk2014giant} optical density $OD$ on excitonic resonances with different principal quantum numbers $n$ obtained for a crystal of length $L=34$ $\mu$m. Reflections off the crystal surface as well as a phonon-assisted background have been accounted for as described in App.~\ref{app:reflections}. The solid lines indicate linear fits used to extract the linear and nonlinear absorption coefficients. 
}
\label{fig3}
\end{figure}

Note that the resonant ($\Delta = 0$) nonlinear absorption coefficient, ${\rm Im}[\chi^{(3)}]$, is always positive and therefore enhances the photon transmission through the material. This is readily understood from the underlying Rydberg exciton blockade mechanism, whereby the strong van der Waals interaction inhibits the simultaneous generation of two nearby excitons and therefore reduces light absorption relative to the linear photon loss determined by $\ell_{\rm abs}^{-1}=-2{\rm Im}[\chi^{(1)}]$. Within the associated Rydberg blockade radius, double excitation of two Rydberg excitons is entirely suppressed, which permits to derive a simple expression, $\chi^{(3)}({\bf 0})=16i\sum_k \frac{g_k^4}{\bar{n}c|\Gamma_k|^2\Gamma_k}$, for the short-distance behavior of the nonlinear response.

Under typical experimental conditions \cite{kazimierczuk2014giant} the light intensity inside the Cu$_2$O material does not vary significantly on the length scale of the blockade radius. In this case we can move the field intensity $|\mathcal{E}(\vec{r}^\prime)|^2$ out of the integral in Eq.(\ref{eq:field_nonlinear}) to obtain a simple analytical solution for the transmitted intensity and the optical depth of the material $ OD = L/\ell_{\rm abs} - \sigma I_0$, where the nonlinear absorption coefficient is given by
\begin{equation} \label{eq:sigma}
\begin{aligned}
\sigma= \frac{2\ell_{\rm abs}\bar{n}(1-{\rm e}^{-L/\ell_{\rm abs}})}{\hbar \omega c}\int {\rm Im}\left[\chi^{(3)}(\mathbf{r})\right]  {\rm d} \vec{r},
\end{aligned}
\end{equation}
determining the dependence of the absorption on the input intensity $I_0$.
This result can be compared directly with the observed response of Cu$_2$O Rydberg excitons by analyzing the absorption spectra measured in \cite{kazimierczuk2014giant}. By properly accounting for known linear effects such as reflection off the material surfaces and phonon-induced background absorption (see App.~\ref{app:reflections}) one can determine the $OD$ generated by each Rydberg-exciton resonance from the measured series of absorption peaks. The obtained optical depth is shown in Fig.\ref{fig3} as a function of the input intensity for different principal quantum numbers of the respective Rydberg state. By fitting a linear intensity dependence to the depicted curves one can extract experimental values for the nonlinear coefficient $\sigma$, which is compared to the theoretical prediction of Eq.~(\ref{eq:sigma}) in Fig.\ref{fig1}. Indeed, the developed theory yields remarkably good agreement for principal quantum numbers $n \lesssim 15$ without any free fitting parameters and indicates a strong enhancement of the nonlinearity with increasing level of Rydberg excitation. At higher values of $n$, however, one finds that the observed nonlinear response even exceeds the theoretical expectation considerably.

We attribute this effect to plasma screening. 
The action of an electron-hole plasma on the excitonic series has been analyzed in recent measurements \cite{heckoetter2018}, where it was found that even rather dilute plasmas can have sizable effects on highly excited Rydberg-exciton states. By controlling the plasma density via laser excitation these experiments demonstrated the consecutive extinction of Rydberg states due to band edge lowering with increasing plasma density. 
In the absence of additional plasma generation, the experiment reaches a maximum principal Rydberg-state quantum number of $n=25$, which can indeed be caused by band-edge lowering due to a residual plasma with a very small density below $1/\mu{\rm m}^3$.

\begin{figure}[t!]
\begin{center}
 \includegraphics[height=.55\textwidth]{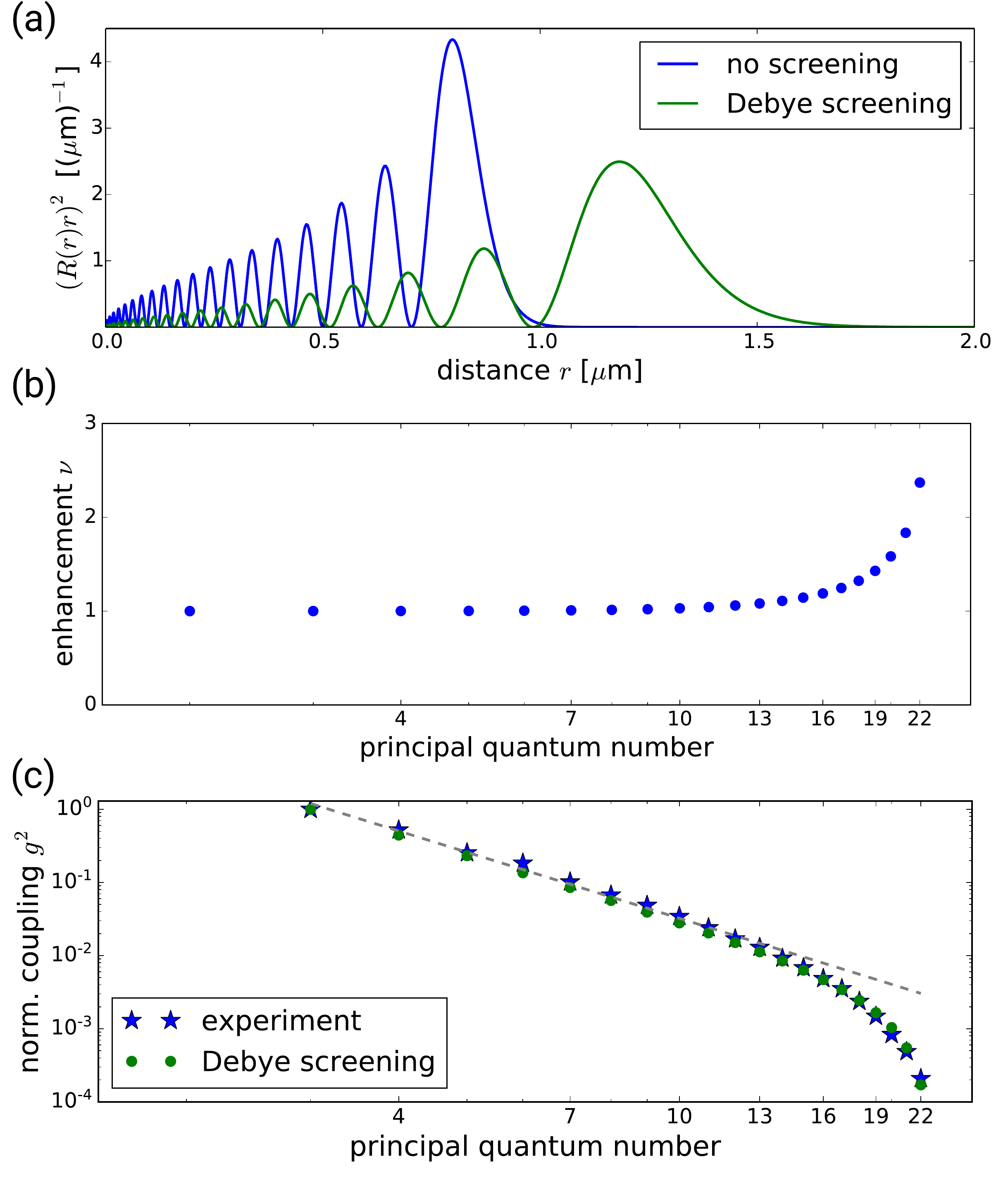}
\end{center}
\caption{Debye screening by a dilute residual plasma in the semiconductor tends to weaken the electron-hole attraction that binds the Rydberg exciton. (a) This effect enhances the extent of the excitonic wave function, as shown exemplarily for the $20p$ state. The corresponding enhancement factor $\nu=\langle r\rangle_{\rm D}/\langle r\rangle_0$ is shown in panel (b) as a function of the principal Rydberg-state quantum number. The depicted growth of the orbital size of the exciton implies a drop of its oscillator strength. As shown in panel (c), the correspondingly calculated oscillator strength (App.~\ref{app:plasma}), is in excellent agreement with the experimentally observed drop \cite{kazimierczuk2014giant} of the light-matter coupling strength $g$. All calculations have been performed for a Debye screening length $\lambda_{\rm D}=0.4 \ \mu$m.}
\label{fig4}
\end{figure}

While such densities are far too low to have any observable effect on typical excitons in their ground states they will eventually affect the properties of highly lying Rydberg exciton states. 
At such low plasma densities $\rho$ and a typical temperature of $T=1.2$ K \cite{kazimierczuk2014giant,heckoetter2018}, we can use Debye theory to describe the exponential screening of the electron-hole interaction that binds the exciton with a screening length $\lambda_{\rm D}=\sqrt{\varepsilon_0\varepsilon_r k_{\rm B}T/(2e^2\rho)}$ \cite{haug2009}. By weakening the exciton binding and removing the long-range character of the electron-hole attraction, this plasma screening limits the observable exciton series, as discussed above and as demonstrated in \cite{kazimierczuk2014giant,heckoetter2018}. As shown in Fig.\ref{fig4}(a), screening tends to spread out the exciton bound-state wave function, an effect which becomes more and more pronounced with increasing principal quantum number as the Rydberg states approach the lowered continuum. 
This is illustrated in Fig.\ref{fig4}(b) where we show the ratio $\nu=\langle r\rangle_{\rm D}/\langle r\rangle_0$ of the calculated exciton radius with ($\langle r\rangle_{\rm D}$) and without ($\langle r\rangle_{0}$) Debye screening. The depicted growth of the exciton size tends to decrease the optical coupling $g_k$ to each exciton, which is determined by the value of the bound-state wave function at small distances $r\rightarrow0$. This suggests a weakening of the \emph{linear} exciton absorption with increasing principal quantum number. This effect is indeed observed experimentally \cite{kazimierczuk2014giant}, and can be quantitatively described by our simple screening model with a screening length of $\lambda_{\rm D}=0.4 \ \mu$m [see Fig.\ref{fig4}(c)].

On the other hand, a growing bound-state wave function naturally increases the strength of the dipole-dipole coupling, $\propto e\langle j|r|k\rangle$, between nearby excitonic Rydberg states $|j\rangle$ and $|k\rangle$ by a factor of $\sim\nu$. As a result, the van der Waals interaction, which scales with the fourth power of the transition dipole moments, is enhanced by a factor of $\nu^4$. An interesting question pertains to screening effects on longer distances, i.e. potential screening of the transition dipoles generated by two interacting excitons. Here, one should note, however, that the rapidly oscillating induced dipoles of the excitons and the response of the free plasma charges will typically operate on vastly different timescales, which tends to suppress Debye screening of dispersive van der Waals interactions in a plasma \cite{mahanty1976,alastuey2007}. 
Here, we adopt a simplified approach in which we neglect dynamical screening of the rapidly oscillating induced dipoles and account for the enhancement of the excitonic van der Waals interaction by a factor $\nu^4$, calculated for the laser-excited pair of $nP$ Rydberg excitons. As shown in Fig.\ref{fig1}(b), this simple approximation provides a remarkably accurate description of the observed nonlinearity across the measured Rydberg series, giving strong experimental evidence for the assumed screening mechanism as well as the resulting plasma-induced enhancement of optical nonlinearities and exciton interactions described in this work. 

In summary, we have presented a theoretical framework to determine the nonlinear optical response of Rydberg excitons in a semiconductor, which we showed to explain previously measured absorption properties of Cu$_2$O in the terms of the strong van der Waals interactions \cite{walther2018interactions} between highly excited exciton states in this material. The calculations imply that the observed nonlinearities correspond to a Rydberg blockade radius of up to 4.5 $\mu$m, within which the simultaneous generation of two excitons is inhibited by their strong interactions. Such a large interaction range suggests itself as a promising platform for manipulating and controlling mesoscopic correlated states of semiconductor excitons. For example, the recently demonstrated fabrication of Cu$_2$O microcrystals \cite{steinhauer2019} can reach dimensions well below this blockade radius, and may thereby make it possible to exploit the enhanced photon coupling provided by a single collective Rydberg excitation in a fully blockaded material.  

Our calculations suggest that the experimentally observed absorption spectrum of Cu$_2$O indicates the presence of a residual low-density plasma, which we found to further enhance the van Waals interaction between Rydberg excitons. Accounting for this plasma-enhancement within a simple screening model yields a remarkably good description of the measured nonlinearities across all observed principal quantum numbers of the excitonic Rydberg states. 
While our simplified treatment therefore provides a promising starting point, it also motivates future work to elucidate the importance of plasma effects on the excitonic energy levels and their wavefunctions as well as the physics of dynamical screening of intra- and inter-exciton interactions on all relevant scales. Clearly, Rydberg states can act as a sensitive probe of their environment and future experiments on their nonlinear optical response could thereby improve our understanding of the non-equilibrium dynamics of interacting excitons, free plasma charges, and their thermalization, e.g., through time-resolved spectroscopic measurements. 

While we have focussed here on absorptive nonlinearities, the integration of Cu$_2$O into optical microcavities, to enhance the relatively weak optical dipole coupling of the $p$-series, may open the door to studying and exploiting giant dispersive nonlinearities in solid-state interfaces. 

\textit{Acknowledgments: }
We thank Hossein Sadeghpour, Julian Heck\"{o}tter, Nikola \v{S}ibali\'c, Jovica Stanojevic, and Stefan Scheel for useful discussions and Manfred Bayer for sharing the experimental data from \cite{kazimierczuk2014giant}. This work has been supported by the EU
through the H2020-FETOPEN Grant No. 800942640378 (ErBeStA), by the DFG through the SPP1929, by the Carlsberg Foundation through the Semper Ardens Research Project QCooL, and by the DNRF through a Niels Bohr Professorship to T. P.

\clearpage
\onecolumngrid
\appendix

\section{Nonlinear optical response of Rydberg excitons} \label{app:eq_motions}
Here, we describe how to develop and solve the coupled dynamics of excitons interacting via dipole-dipole interactions, as given by the equation of motion (Eq.~\ref{eq:ex_nonlinear})
\begin{equation}\label{eq:supp_eq_boson}
\begin{aligned} 
 \frac{\partial}{\partial t}\hat{X}_k (\vec{r}) &= -ig_k \mathcal{E}(\vec{r}) - \frac{\Gamma_k}{2}\hat{X}_k (\vec{r})
 - i \sum_{iji^\prime} \int d\vec{r}^\prime V_{i^\prime k,ij}(\vec{r}-\vec{r}^\prime) \hat{X}_{i^\prime}^\dagger (\vec{r}^\prime) \hat{X}_{i} (\vec{r}^\prime) \hat{X}_{j} (\vec{r})
\end{aligned}
\end{equation} 
with $\Gamma_k = \gamma_k -2i\Delta_k$ and $V_{ij,kl} = V_{ji,lk} = V_{kl,ij}$ real and where we suppressed the time argument $\hat{X}(\vec{r}) \equiv \hat{X}(\vec{r},t)$. The goal is to obtain a closed analytic solution of $\langle \hat{X}_k (\vec{r}) \rangle$ in the steady state. We first develop the hierarchy of equations of motion for higher-order correlators emerging from Eq.~(\ref{eq:supp_eq_boson}), truncating the three-particle correlations and, thus, closing the system. This captures the lowest-order nonlinearity exactly, which is the dominant term at low densities and driving strengths. 
The system is then given by the following set of equations
\begin{equation}\label{eq:supp_correlator1}
\begin{aligned}
 \frac{\partial}{\partial t} \hat{X}_k^\dagger(\vec{r}^\prime) \hat{X}_j(\vec{r}) &= 
 -ig_j \mathcal{E}(\vec{r}) \hat{X}_k^\dagger (\vec{r}^\prime) + ig^*_k \mathcal{E}^*(\vec{r}^\prime) \hat{X}_j(\vec{r}) 
 - b_j \sum_i \Omega_i \hat{X}_k^\dagger(\vec{r}^\prime) \hat{X}_i (\vec{r})
 - \left( \frac{\Gamma_j}{2} + \frac{\Gamma_k^*}{2}\right) \hat{X}_k^\dagger(\vec{r}^\prime) \hat{X}_j(\vec{r})
\end{aligned}
\end{equation}

\begin{equation} \label{eq:supp_correlator2}
\begin{aligned}
 \frac{\partial}{\partial t} \hat{X}_j (\vec{r}) \hat{X}_l (\vec{r}^\prime) &= 
 -ig_j \mathcal{E}(\vec{r}) \hat{X}_l(\vec{r}^\prime) 
 - \frac{\Gamma_j}{2} \hat{X}_j (\vec{r}) \hat{X}_l (\vec{r}^\prime)
 -ig_l \mathcal{E}(\vec{r}^\prime) \hat{X}_j(\vec{r})
 - \frac{\Gamma_l}{2} \hat{X}_j (\vec{r}) \hat{X}_l (\vec{r}^\prime)
 -i \sum_{ij^\prime} V_{jl,ij^\prime}(\vec{r}-\vec{r}^\prime) \hat{X}_i (\vec{r}) \hat{X}_{j^\prime} (\vec{r}^\prime)
\end{aligned}
\end{equation}

\begin{equation} \label{eq:supp_correlator3}
\begin{aligned}
 \frac{\partial}{\partial t} \hat{X}_i^\dagger(\vec{r}) \hat{X}_j(\vec{r}) \hat{X}_l (\vec{r}^\prime) 
 &= ig^*_i \mathcal{E}^*(\vec{r}) \hat{X}_j(\vec{r})\hat{X}_l(\vec{r^\prime})
 -\frac{\Gamma_i^*}{2} \hat{X}_i^\dagger(\vec{r}) \hat{X}_j(\vec{r}) \hat{X}_l (\vec{r}^\prime)
 -ig_j \mathcal{E}(\vec{r}) \hat{X}_i^\dagger(\vec{r}) \hat{X}_l(\vec{r}^\prime)
 -\frac{\Gamma_j}{2} \hat{X}_i^\dagger(\vec{r}) \hat{X}_j(\vec{r}) \hat{X}_l (\vec{r}^\prime) \\
 &-ig_l \mathcal{E}(\vec{r}^\prime) \hat{X}_i^\dagger(\vec{r}) \hat{X}_j(\vec{r})
 -\frac{\Gamma_l}{2} \hat{X}_i^\dagger(\vec{r}) \hat{X}_j(\vec{r}) \hat{X}_l (\vec{r}^\prime)
 -i \sum_{kj^\prime} V_{jl,kj^\prime}(\vec{r}-\vec{r}^\prime) \hat{X}_i^\dagger(\vec{r})\hat{X}_k(\vec{r}) \hat{X}_{j^\prime}(\vec{r}^\prime)
\end{aligned}
\end{equation}

\begin{equation} \label{eq:supp_correlator4}
\begin{aligned}
 \frac{\partial}{\partial t} \hat{X}^\dagger_i(\vec{r})\hat{X}_j(\vec{r}) 
 &= ig^*_i \mathcal{E}^*(\vec{r}) \hat{X}_j(\vec{r}) 
 - \frac{\Gamma_i^*}{2} \hat{X}^\dagger_i(\vec{r})\hat{X}_j(\vec{r})
 -ig_j \mathcal{E}(\vec{r}) \hat{X}^\dagger_i(\vec{r})
 - \frac{\Gamma_j}{2} \hat{X}^\dagger_i(\vec{r})\hat{X}_j(\vec{r}) \\
 &+i \int d\vec{r}^\prime \sum_{kj^\prime i^\prime} V_{i^\prime i,kj} (\vec{r}-\vec{r}^\prime)\hat{X}_{j^\prime}^\dagger(\vec{r})\hat{X}_{k}^\dagger(\vec{r}^\prime)\hat{X}_{i^\prime}(\vec{r}^\prime)\hat{X}_{j}(\vec{r}) \\
 &-i \int d\vec{r}^\prime \sum_{kj^\prime i^\prime} V_{i^\prime j,kj^\prime} (\vec{r}-\vec{r}^\prime)\hat{X}_{i}^\dagger(\vec{r})\hat{X}_{i^\prime}^\dagger(\vec{r}^\prime)\hat{X}_{k}(\vec{r}^\prime)\hat{X}_{j^\prime}(\vec{r}).
\end{aligned}
\end{equation}

The bosonic operators are defined in a coordinate system aligned with the laser propagation (polarization) for circularly (linearly) polarized light (``lab frame''). For the interacting case, however, it is convenient to treat the problem in a coordinate system aligned with the axis joining pairs of excitons (``molecular frame''). The angular dependence of the Rydberg interaction is then transferred into single-exciton coupling elements that are easily evaluated. The rotation of excitonic states is given by Wigner-d functions 
\begin{align}
 \hat{X}_i^{\text{lab}} = \sum_{k}d^*_{i,k}(\theta)\hat{X}_k^{\text{mol}} \qquad \hat{X}_l^{\text{mol}} = \sum_{i}d_{i,l}(\theta)\hat{X}_i^{\text{lab}}\\
 \sum_{k} d^*_{i,k}(\theta)d_{j,k}(\theta) = \delta_{ij} \qquad \sum_{i} d^*_{i,k}(\theta)d_{i,l}(\theta) = \delta_{kl},
\end{align}
where we note that a rotation transformation only mixes the $m$ submanifold, not quantum numbers $l$ and $n$, such that all $\Gamma_{k}$ remain unchanged. Transforming the exciton operators from Eqs.~(\ref{eq:supp_correlator1})-(\ref{eq:supp_correlator4}) into the molecular frame, we find that the form of the equations is unchanged if we introduce the substitution 
\begin{align} \label{eq:trafo_g}
 g_k \to \sum_{i}d_{i,k} (\theta) g_i.
\end{align}
The angle $\theta \in [0,\pi ]$ is referred to as the interaction angle \cite{weber2017} and is defined as the angle between the lab and molecular $\hat{\mathbf{z}}$-axes. The simple transformation shows mathematically that the angular dependence can fully be absorbed into the coupling strength $g(\theta)$, while the interaction now is a function of the exciton separation only $V_{ij,i^\prime j^\prime}(\vec{r}-\vec{r}^\prime) \to V_{ij,i^\prime j^\prime}(|\vec{r}-\vec{r}^\prime|)$. We drop the argument for the molecular frame in the following until the distinction becomes essential in Eq.~(\ref{eq:supp_eq_boson_new_basis}).
Next, we define the interaction matrix
\begin{align}
 \mathcal{V}_{ij,kl} = V_{ij,kl} - \left( \Delta_i + \Delta_j \right) \delta_{ik}\delta_{jl},
\end{align}
corresponding to the matrix which is diagonalized numerically \cite{walther2018interactions}. 
The matrix $\mathcal{V}$ can be identified in each of the equations containing the interaction. Explicitly, we re-write from Eq.~(\ref{eq:supp_correlator2}) and Eq.~(\ref{eq:supp_correlator3})
\begin{equation}
\begin{aligned}
 &- \frac{\Gamma_j}{2} \hat{X}_j (\vec{r}) \hat{X}_l (\vec{r}^\prime) 
 - \frac{\Gamma_l}{2} \hat{X}_j (\vec{r}) \hat{X}_l (\vec{r}^\prime)
 -i \sum_{ij^\prime} V_{jl,ij^\prime}(|\vec{r}-\vec{r}^\prime|) \hat{X}_i (\vec{r}) \hat{X}_{j^\prime} (\vec{r}^\prime) \\
 = &- \left( \frac{\gamma_j}{2} + \frac{\gamma_l}{2} \right) \hat{X}_j (\vec{r}) \hat{X}_l (\vec{r}^\prime) 
 -i \sum_{ij^\prime} \mathcal{V}_{jl,ij^\prime}(|\vec{r}-\vec{r}^\prime|) \hat{X}_i (\vec{r}) \hat{X}_{j^\prime} (\vec{r}^\prime)
\end{aligned}
\end{equation}
We, therefore, pursue a closed solution in the basis where $\mathcal{V}$ is diagonal. This basis has been obtained numerically \cite{walther2018interactions} and its transformation can formally be summarized as 
\begin{align}
 \Ket{\mu} = \sum_{ij} c^*_{ij, \mu} \Ket{ij} \qquad \Ket{ij} = \sum_{\mu} c_{ij, \mu} \Ket{\mu} \\
 \sum_{ij} c^*_{ij, \mu} c_{ij, \mu^\prime} = \delta_{\mu \mu^\prime} \qquad \sum_{\mu} c^*_{(ij), \mu} c_{(ij)^\prime, \mu} = \delta_{(ij)(ij)^\prime},
\end{align}
where $|\mu\rangle$ are the eigenstates of $\mathcal{V}$ with the corresponding eigenvalues $U_\mu$. Because of the Hamilitonian's symmetries the eigenstates are invariant under particle exchange $c_{ij,\mu} = c_{ji,\mu}$. We define the two-exciton operators (here $g$ denotes the vacuum ground state)
\begin{align}
 \hat{Z}_{gg,\mu}(\vec{r},\vec{r}^\prime) &\equiv \sum_{jl} c_{jl,\mu} \hat{X}_j(\vec{r})\hat{X}_l(\vec{r}^\prime)\\
 \hat{Y}_{ig,\mu}(\vec{r},\vec{r}^\prime) &\equiv \sum_{jl} c_{jl,\mu} \hat{X}_i^\dagger(\vec{r}) \hat{X}_j(\vec{r}) \hat{X}_l(\vec{r}^\prime) = \hat{X}_i^\dagger(\vec{r}) \hat{Z}_{gg,\mu}(\vec{r},\vec{r}^\prime) .
\end{align}
Transforming Eq.~(\ref{eq:supp_correlator2}) and Eq.~(\ref{eq:supp_correlator3}) to the new basis gives
\begin{equation}
\begin{aligned}
 &\frac{\partial}{\partial t} \hat{Z}_{gg,\mu}(\vec{r},\vec{r}^\prime)
 =\sum_{jl} c_{jl,\mu} \left( -ig_j \mathcal{E}(\vec{r}) \hat{X}_l(\vec{r}^\prime) -ig_l \mathcal{E}(\vec{r}^\prime) \hat{X}_j(\vec{r}) \right)
 - \sum_{\bar{\mu}}\bar{\gamma}_{\mu,\bar{\mu}} \hat{Z}_{gg,\bar{\mu}}(\vec{r},\vec{r}^\prime) 
 - i U_\mu (|\vec{r}-\vec{r}^\prime|) \hat{Z}_{gg,\mu}(\vec{r},\vec{r}^\prime)
\end{aligned}
\end{equation}

\begin{equation}
\begin{aligned}
\frac{\partial}{\partial t} \hat{Y}_{ig,\mu}(\vec{r},\vec{r}^\prime)
 &= ig^*_i \mathcal{E}^*(\vec{r}) \hat{Z}_{gg,\mu}(\vec{r},\vec{r}^\prime) 
 + \sum_{jl} c_{jl,\mu} \left( -ig_j\mathcal{E}(\vec{r}) \hat{X}_i^\dagger(\vec{r})\hat{X}_l(\vec{r}^\prime) 
 -ig_l \mathcal{E}(\vec{r}^\prime) \hat{X}_i^\dagger(\vec{r})\hat{X}_j(\vec{r}) \right) \\
 &-\frac{\Gamma_i^*}{2} \hat{Y}_{ig,\mu}(\vec{r},\vec{r}^\prime) 
 - \sum_{\bar{\mu}}\bar{\gamma}_{\mu,\bar{\mu}} \hat{Y}_{ig,\bar{\mu}}(\vec{r},\vec{r}^\prime)
 - i U_\mu (|\vec{r}-\vec{r}^\prime|) \hat{Y}_{ig,\mu}(\vec{r},\vec{r}^\prime),
\end{aligned}
\end{equation}
where we defined the auxiliary matrix
\begin{align}
 \bar{\gamma}_{\mu,\bar{\mu}} &= \sum_{jl} c_{jl,\mu} \gamma_j c^*_{jl,\bar{\mu}}.
\end{align}
We have now exposed a form of the equations that is diagonal in the interaction. The off-diagonal components are only contained in $\bar{\gamma}_{\mu,\bar{\mu}}$. If these terms are diagonal, all equations can be solved exactly. Here, we neglect the off-diagonal contributions from these terms, an approximation that is justified by the slowly-varying values $\gamma_j$ in the Rydberg manifold. Then, taking expectation values and considering the steady state, we find the diagonal solutions
\begin{equation}
\begin{aligned}
 \langle \hat{Z}_{gg,\mu}(\vec{r},\vec{r}^\prime) \rangle \approx
 \frac{\sum_{jl} c_{jl,\mu} \left( -ig_j \mathcal{E}(\vec{r}) \langle \hat{X}_l(\vec{r}^\prime)\rangle -ig_l \mathcal{E}(\vec{r}^\prime) \langle \hat{X}_j(\vec{r}) \rangle \right)}
 {\bar{\gamma}_{\mu,\mu}
 + i U_\mu (|\vec{r}-\vec{r}^\prime|)}
\end{aligned}
\end{equation}
\begin{equation}\label{eq:supp_Y}
\begin{aligned}
 &\langle \hat{Y}_{ig,\mu}(\vec{r},\vec{r}^\prime) \rangle 
 \approx \frac{g^*_i \mathcal{E}^*(\vec{r}) \langle\hat{Z}_{gg,\mu}(\vec{r},\vec{r}^\prime)\rangle 
 + \sum_{jl}c_{jl,\mu} \left( -ig_j\mathcal{E}(\vec{r}) \langle \hat{X}_i^\dagger(\vec{r})\hat{X}_l(\vec{r}^\prime) \rangle
 -ig_l \mathcal{E}(\vec{r}^\prime) \langle \hat{X}_i^\dagger(\vec{r})\hat{X}_j(\vec{r}) \rangle\right)}
 {\frac{\Gamma_i^*}{2} 
 +\bar{\gamma}_{\mu,\mu} 
 + i U_\mu (|\vec{r}-\vec{r}^\prime|) }.
\end{aligned}
\end{equation}
A diagonal solution can also be found from Eq.~(\ref{eq:supp_correlator1})
\begin{equation}
\begin{aligned}
 \langle \hat{X}_k^\dagger(\vec{r}^\prime) \hat{X}_j(\vec{r}) \rangle \approx \frac{-ig_j \mathcal{E}(\vec{r}) \langle \hat{X}_k^\dagger (\vec{r}^\prime)\rangle + ig^*_k \mathcal{E}^*(\vec{r}^\prime) \langle \hat{X}_j(\vec{r}) \rangle}{\left( \frac{\Gamma_j}{2} + \frac{\Gamma_k^*}{2}\right) }.
\end{aligned}
\end{equation}
Eq.~(\ref{eq:supp_correlator4}) takes a special position as it contains fourth-order terms in the exciton operators. As can be seen from the perturbation expansion in the field $\mathcal{E}$, these do not contribute to next-to-leading (i.e. third) order and can be dropped. We obtain for the diagonal solution
\begin{align}
 \langle \hat{X}^\dagger_i(\vec{r})\hat{X}_j(\vec{r}) \rangle \approx \frac{ig^*_i \mathcal{E}^*(\vec{r}) \langle \hat{X}_j (\vec{r})\rangle -ig_j \mathcal{E}(\vec{r}) \langle \hat{X}^\dagger_i(\vec{r})\rangle}{\frac{\Gamma_i^*}{2}+\frac{\Gamma_j}{2}}.
\end{align}
Finally, we can rewrite the convolution from Eq.~(\ref{eq:supp_eq_boson}) using the same basis in the steady state
\begin{equation}\label{eq:supp_eq_boson_new_basis}
\begin{aligned} 
\langle \hat{X}_k^\text{lab} (\vec{r}) \rangle \approx &-\frac{2ig_k^\text{lab}}{\Gamma_k} \mathcal{E}(\vec{r})
- \frac{2i}{\Gamma_k} \int d\vec{r}^\prime \sum_\mu \sum_{i^\prime} \sum_{k^\prime} \left( U_{\mu}(|\vec{r}-\vec{r}^\prime|) +\Delta_{i^\prime} + \Delta_{k^\prime} \right) c^*_{i^\prime k^\prime, \mu} d_{k,k^\prime}(\theta) \langle \hat{Y}_{g i^\prime, \mu}(\vec{r},\vec{r}^\prime) \rangle^\text{mol} .
\end{aligned}
\end{equation} 
Eq.~(\ref{eq:supp_eq_boson_new_basis}) is an implicit equation for the desired expectation value $\langle \hat{X}_k \rangle$. We solve it in a perturbative expansion in the field $\mathcal{E}$ (weak-field expansion) that corresponds to a cluster expansion in the excitonic correlators. The resulting expression is of first- and third order in the field and is substituted into Eq.~(\ref{eq:field}) to give the nonlinear light propagation equation
\begin{align} \label{eq:nonlinear_phot_prop}
 i\frac{\partial}{\partial z} \mathcal{E} (\vec{r}) = \chi^{(1)} \mathcal{E} (\vec{r}) + \int d\vec{r}^\prime \chi^{(3)}(|\vec{r}-\vec{r}^\prime|, \theta) |\mathcal{E}(\vec{r}^\prime)|^2 \mathcal{E}(\vec{r}).
\end{align}
with the linear susceptibility
\begin{align}
 \chi^{(1)} = - \sum_n \frac{2ig_n^2}{\bar{n}c \Gamma_n}
\end{align}
and the nonlinear susceptibility
\begin{align} \label{eq:chi3}
 \chi^{(3)}(|\vec{r}-\vec{r}^\prime|, \theta) = -\frac{8}{\bar{n}c} \sum_\mu D_\mu(\theta) \frac{B_\mu(\theta) U_\mu(|\vec{r}-\vec{r}^\prime|) + C_\mu(\theta)}{\bar{\gamma}_\mu + iU_\mu (|\vec{r}-\vec{r}^\prime|)}
\end{align}
with the auxiliary variables 
\begin{align}
 D_\mu(\theta) &= \sum_{jl} c_{jl,\mu} G_{jl}(\theta) \left[ \frac{1}{\Gamma_l} + \frac{1}{\Gamma_j} \right] \\
 B_\mu(\theta) &= \sum_{i^\prime k} c_{i^\prime k, \mu}^* \frac{G_{i^\prime k}(\theta)}{\Gamma_{i^\prime}^*\Gamma_{k}} \\
 C_\mu(\theta) &= \sum_{i^\prime k} \left( \Delta_{i^\prime} + \Delta_k \right) c_{i^\prime k,\mu}^* \frac{G_{i^\prime k}(\theta)}{\Gamma_{i^\prime}^*\Gamma_{k}}\\
 G_{xy}(\theta) &= \sum_{pq} g_p g_q d_{p,x}(\theta)d_{q,y}(\theta).
\end{align}
These are the explicit expressions for the susceptibilities given in Eq.(\ref{eq:field_nonlinear}).

\section{Extracting nonlinearities from data} \label{app:reflections}
In this section, we describe how we obtain the optical nonlinearity shown in Fig.\ref{fig3} from experimental data by accounting for reflections off the crystal surface and the well-known phonon-assisted absorption background underlying the excitonic Rydberg series in Cu$_2$O. This procedure involves extracting the single-exciton dipole elements $g_n$ for each exciton and calculating the effective light intensity entering the crystal. The former is done in the linear regime where transmission is given via the total transmission coefficient of the crystal $t_c$ and the transmission coefficient at the first (second) surface $t_1$ ($t_2$)
\begin{align} \label{eq:trans_lin_exact}
 I^\text{out} = I^\text{in} t_1 t_c t_2 \sum_{k=0}^{\infty} \left[ (1-t_1)(1-t_2)t^2_c \right]^k .
\end{align}
For the experimental conditions of the light beam traversing orthogonally through the crystal surrounded by a Helium bath ($n_\text{He} \approx 1$), we find $t_1=t_2=4 \bar{n}/ (1 + \bar{n})^2 \approx 0.78$. Since $t_c < 1$, the $(k>0)$-terms from Eq.~(\ref{eq:trans_lin_exact}) can be dropped at an error $<5\%$. We further note that the transmission coefficient in the crystal is the product of the phonon background and the excitonic contribution $t_c = t_Xt_{bg}$. The total linear optical density of the crystal at incident frequency $\omega$ is then given by
\begin{align}
 OD_\text{tot}(\omega) = -\ln \left( \frac{I^\text{out}(\omega)}{I^\text{in}} \right) = -\ln \left(t_1 t_X(\omega)t_{bg}(\omega) t_2 \right) = OD_\text{add}(\omega) - \ln \left( t_X(\omega) \right).
\end{align}
The frequency dependence in $OD_\text{add}(\omega)$ originates from the phonon-assisted background and varies slowly across each exciton resonance. We therefore approximate it by a linear function for each resonance and extract the excitonic dipole couplings from $OD(\omega)=OD_\text{tot}(\omega) - OD_\text{add}(\omega)$, a function that vanishes in between the exciton resonances. Note that this procedure exactly eliminates reflection effects in linear transmission.

In the nonlinear regime, however, the absolute input intensity is important. For a general application of our results, the main text discusses transmission inside the crystal, i.e. independent of setup-dependent reflection properties. To obtain a comparison with the experimental values we, therefore, use as input intensity $I_0 = t_1 I^\text{in}$ to extract the nonlinearities (Fig.\ref{fig1}b).

\section{Plasma screening} \label{app:plasma}
\begin{figure}[h!]
 \begin{center}
 \includegraphics[height=.37\textwidth]{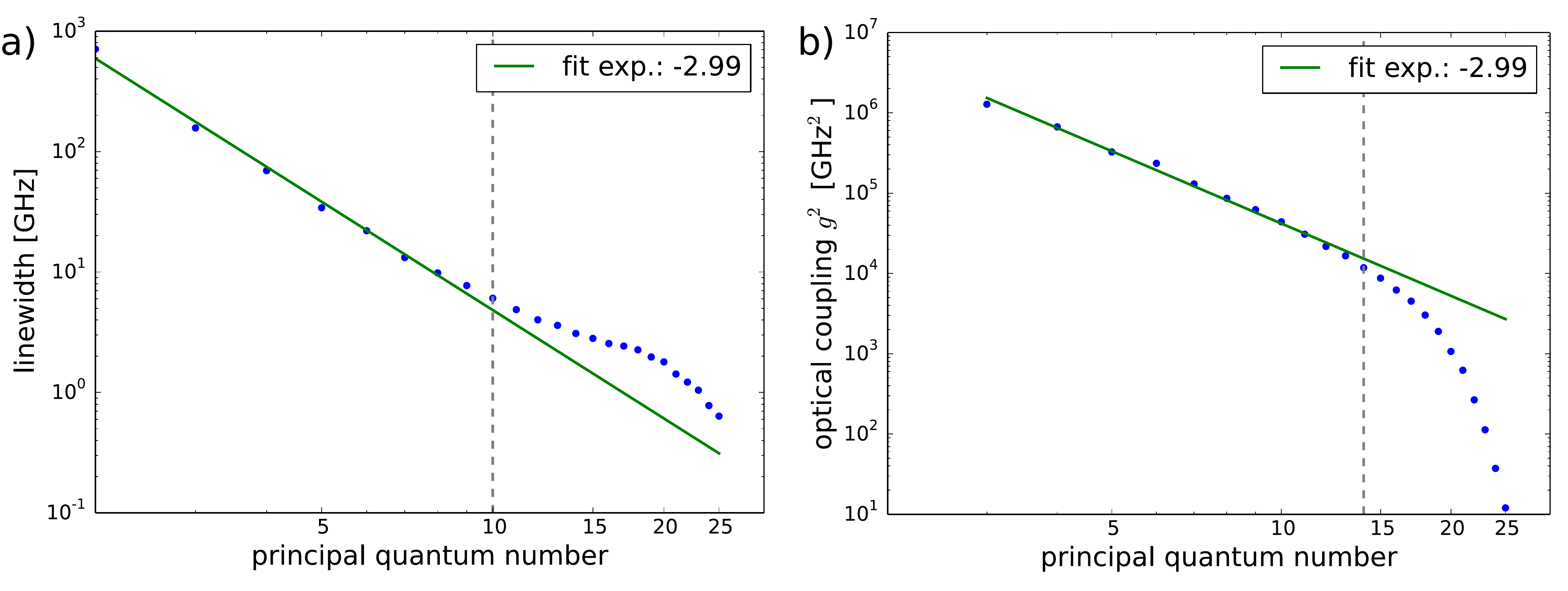}
 \end{center}
  \caption{\textbf{Comparison between experiment and radiative scaling model.} a) Linewidth versus principal quantum number. The data (taken from \cite{kazimierczuk2014giant}) is fit well by a power law $\gamma \propto n^{-3}$ for small $n$, then stronger decay/dephasing sets in. b) Optical coupling $g^2$ (proportional to the square of the dipole moment) versus principal quantum number. Fit $g^2 \propto n^{-3}$ accurate at small $n$, but deviates strongly at large $n$. The dashed vertical lines mark the right end of the fit region.}
\label{fig:scaling}
\end{figure}

The Rydberg series in Cu$_2$O strongly resembles that of hydrogen, as the scaling of the energies ($\propto n^{-2}$) and the absorption linewidths ($\propto n^{-3}$) is consistent with expectations based on a hydrogenic model of the exciton. However, closer inspection shows that the linewidths for $n \gtrsim 10$ deviate from the mentioned scaling are actually much broader. Moreover, the highest resonances seem to sink into a background, such that states with $n > 25$ cannot be resolved. 
The linearized Maxwell-Bloch equations (main text) give the peak height on resonance in terms of the optical depth 
\begin{align} \label{eq:OD}
 OD = -\ln(I(L)/I(0)) = \frac{4g^2L}{\bar{n}c\gamma}. 
\end{align}
Here, $c$ is the speed of light, $\bar{n}$ is the refractive index, $L$ is the sample length, $g$ is the optical coupling rate and $\gamma$ is the phenomenological exciton decay rate accounting for all excitonic loss and dephasing mechanisms. In atomic systems, the crudest scaling arguments suggest that $g \propto n^{-3/2}$ and $\gamma \propto n^{-3}$, making the Rydberg series a succession of ever narrower peaks of equal height. 
The shrinking and final disappearance of the Rydberg states in Cu$_2$O can, therefore, quite generally be explained as an ``imbalanced'' scaling of the dipole coupling and the decay rate. We now look at the experimental data \cite{kazimierczuk2014giant} to separate these effects (Fig.\ref{fig:scaling}).
It turns out that the decay/decoherence mechanisms, for small quantum numbers, scale roughly as would be expected from radiative decay. Based on the scaling picture of the wave function, it is reasonable to expect that even the phononic contribution would show the same scaling in the Rydberg manifold \cite{toyozawa1958, toyozawa1959, toyozawa1964}. At larger $n$, the lines get significantly broader than expected from scaling arguments. Although the assumptions about possible decoherence mechanisms are kept at a minimum here, they cannot alone account for the disappearing peaks. This can be seen by extracting the effective coupling strength of each exciton transition from the experimental peak height and comparing with Eq.~(\ref{eq:OD}) (Fig.\ref{fig:scaling}b). Beyond $n \approx 14$, the optical coupling strength plummets, falling much faster than $g^2 \propto n^{-3}$. Although similar effects can be seen in Rydberg blockade \cite{sevincli2011}, this origin can be ruled out in the ultralow-intensity limit. Thus, this observation indicates a coupling mechanism that drastically changes the states themselves. 

\begin{figure}[h!]
 \begin{center}
 \includegraphics[height=.37\textwidth]{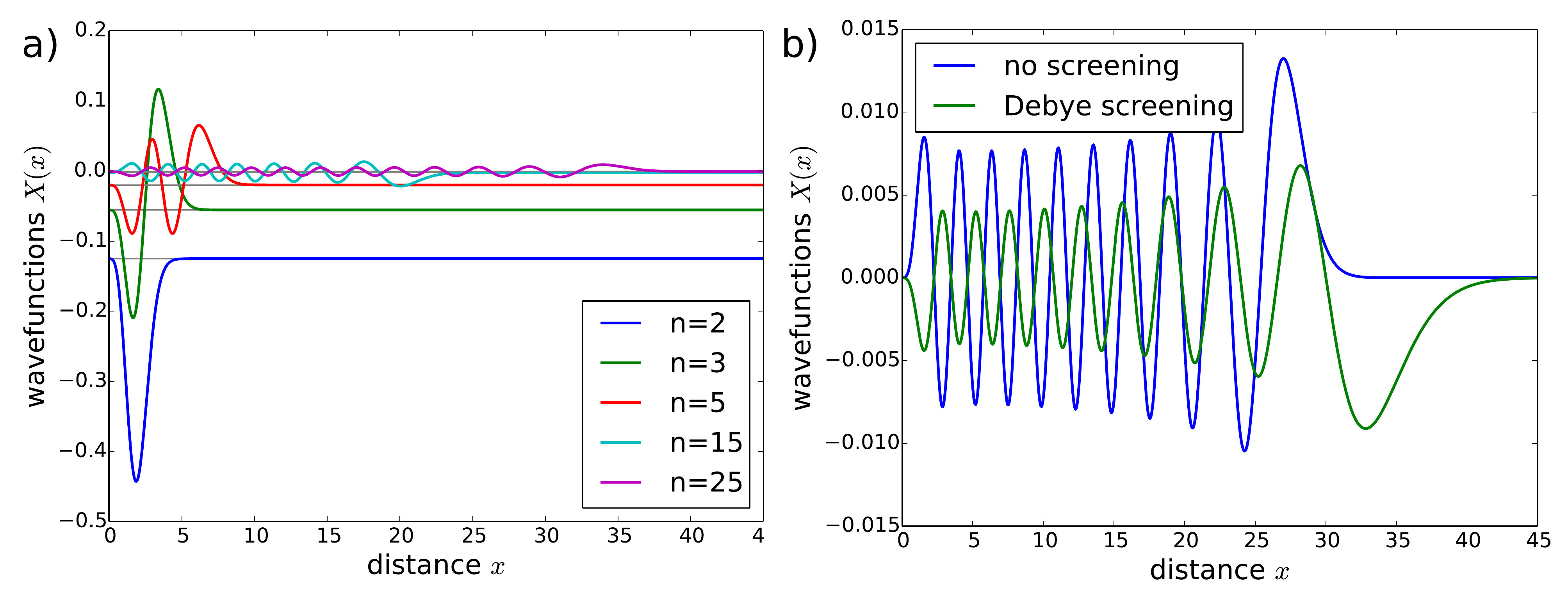}
 \end{center}
  \caption{\textbf{Wave functions from Debye model.} a) Examples of the Rydberg $p$-series without screening in scaled coordinates. The wave functions are offset by their energies $\bar{E}$, marked in gray horizontal lines. b) Comparison of the $15p$ states without screening (blue) and with screening (green, $\bar{\kappa} = 0.004$). The plasma lifts the energy towards the continuum and the pushes the exciton orbit outwards.}
\label{fig:example}
\end{figure}

A likely cause for the observed loss of oscillator strength is so-called plasma screening, which can be due to induced free carriers \cite{heckoetter2018} or due to residual plasma from, e.g., impurities. In the low-density limit, a plasma is non-degenerate and can be described by classical Debye theory \cite{klingshirn2012}. The effect of screening is a simple renormalization into a Yukawa-type potential
\begin{align}\label{eq:screened_pot}
 V(r) = \frac{1}{4\pi \varepsilon_0 \varepsilon_r} \frac{1}{r} \rightarrow V(r) = \frac{1}{4\pi \varepsilon_0 \varepsilon_r} \frac{1}{r} \exp(-\kappa r),
\end{align}
where the inverse screening length $\kappa = \lambda^{-1}_{\rm D}$ is given by
\begin{align}
 \kappa = \sqrt{\frac{2\rho_\text{plasma} e^2}{\varepsilon_0 \varepsilon_r k_b T}}
\end{align}
with the plasma density $\rho_\text{plasma}$ and the plasma temperature $T$, which is typically assumed to be the semiconductor temperature \cite{heckoetter2018}. To check if plasma screening gives the observed features, especially a drop of the oscillator strength at large $n$, we solve the radial Schr\"{o}dinger equation
\begin{align}
 -\frac{\hbar^2}{2 \mu} \left[ \frac{d^2}{dr^2} + \frac{2}{r} \frac{d}{dr} \right] R(r) + \left[ \frac{\hbar^2}{2 \mu} \frac{l(l+1)}{r^2} + V(r) \right] R(r) = E R(r)
\end{align}
with the reduced mass $\mu = m_e m_h /(m_e + m_h)$. Scaling the equation ($r=\bar{r} \cdot \frac{4\pi \varepsilon_0\varepsilon_r\hbar^2}{\mu e^2}$ and $E=\bar{E} \cdot \frac{\mu e^4}{\hbar^2 (4\pi \varepsilon_0\varepsilon_r)^2}$) leaves a single free parameter $\bar{\kappa} = \frac{4\pi \hbar^2}{\mu e} \sqrt{(2\rho_\text{plasma} \varepsilon_0 \varepsilon_r)/(k_b T)}$
\begin{align}
 -\frac{1}{2} \left[ \frac{d^2}{dr^2} + \frac{2}{r} \frac{d}{dr} \right] R(r) + \left[ \frac{l(l+1)}{2r^2} + \frac{1}{r}e^{-\bar{\kappa} r} \right] R(r) = E R(r),
\end{align}
where the overhead bars were dropped for simplicity. Anticipating the structure of hydrogenic wave functions, we follow the numerical procedure in \cite{bhatti1981} to re-scale $x = \sqrt{r}$ and $R(r) = X(r)r^{-3/4}$. We numerically solve the resulting equation
\begin{align}
 -\frac{1}{8x^3}\frac{d^2}{dx^2} X(x) + \left[ \frac{3}{32} \frac{1}{x^4} + \frac{l(l+1)}{2x^4} + \frac{1}{x^2}e^{-\bar{\kappa} x^2} \right] X(x) = E X(x)
\end{align}
in a finite difference-method diagonalization (Fig.\ref{fig:example}). 
As is well-known, the screening at large distances leaves only a finite number of bound states from the Coulomb potential. Even more relevant to the question of disappearing lines in Cu$_2$O, though, is how the oscillator strength of each of the Rydberg lines changes with the plasma density. The optical transition elements of ``forbidden'' transitions, as is the case of excitons in Cu$_2$O, is proportional to $|\frac{dR(0)}{dr}|^2$, making the $p$-series active \cite{elliott1957}. Numerically, we evaluate
\begin{align}
 g^2 \propto \left|\frac{dR(0)}{dr} \right|^2 = \lim_{x \rightarrow 0} \left| \frac{1}{2 x^{5/2}} \frac{dX(x)}{dx} - \frac{3}{4 x^{7/2}} X(x) \right|^2.
\end{align}
\begin{figure}[h!]
 \begin{center}
 \includegraphics[height=.4\textwidth]{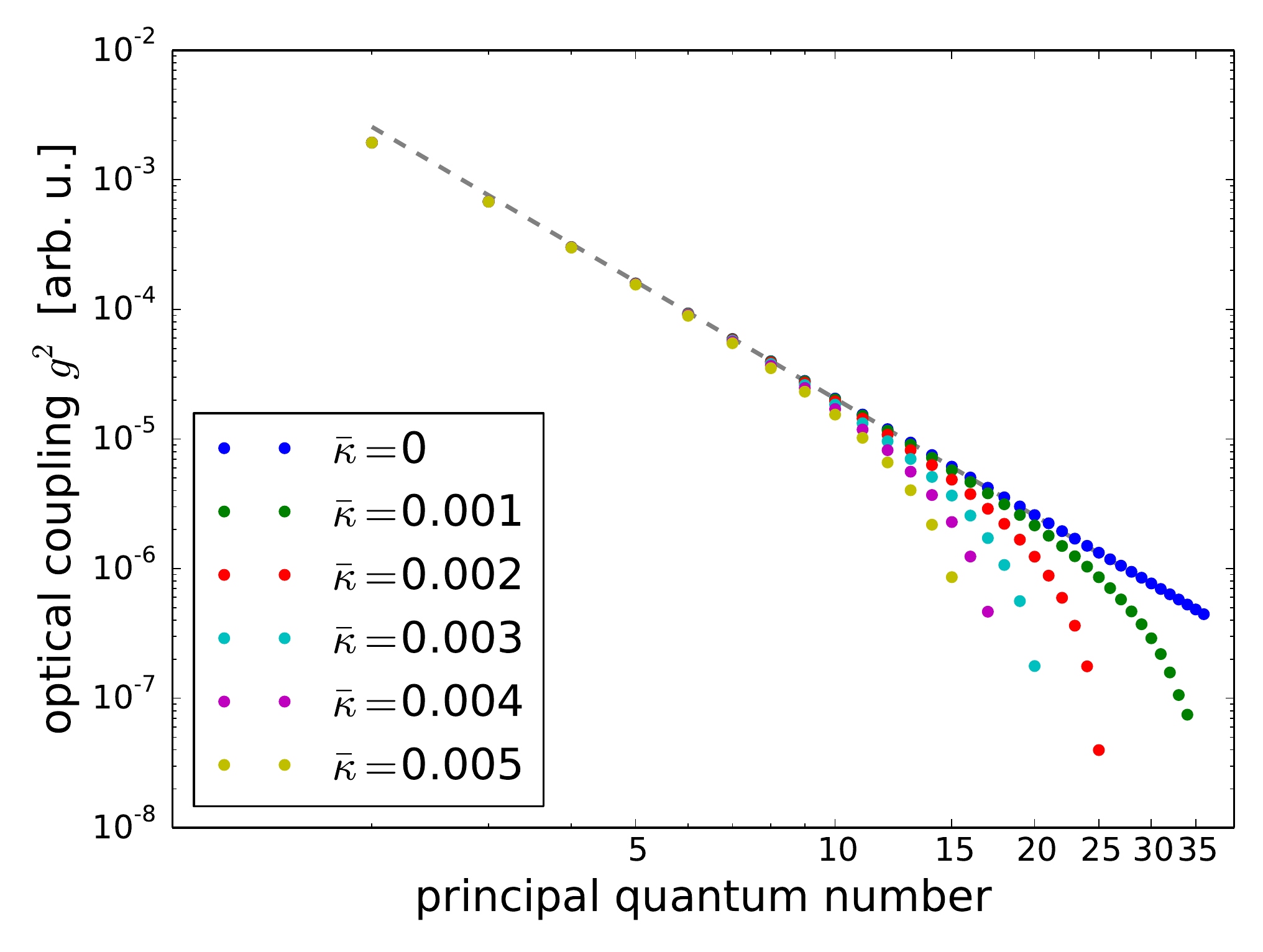}
 \end{center}
  \caption{\textbf{Screened oscillator strengths.} Oscillator strengths $g^2$ for bound states in the screened potential of Eq.~(\ref{eq:screened_pot}) for various scaled inverse screening lengths $\bar{\kappa}$. The gray dashed lines indicates the expected scaling $n^{-3}$ of the unscreened $p$-series in hydrogen. }
\label{fig:oscillator_screened}
\end{figure}
The result (Fig.\ref{fig:oscillator_screened}) shows that before a bound state is pushed into the continuum it loses oscillator strength, qualitatively reproducing the effect observed in the spectra (Fig.\ref{fig:scaling}b). We can estimate that the $n=25$ state and its oscillator strength disappear when $\bar{\kappa} \approx 2\cdot 10^{-3}$. A more accurate numerical optimization of the relative deviation renders $\bar{\kappa} = 2.5\cdot 10^{-3}$. Putting in experimental parameters $m_e = 0.985m_0$, $m_h = 0.575 m_0$ ($m_0$ is the free electron mass) and $\varepsilon_r = 7.5$ for a temperature of $T=1.2$~K  \cite{kazimierczuk2014giant} gives a plasma density of well below $\frac{1}{(\mu \text{m})^3}$ and a screening length of
\begin{align}
 \lambda_\text{D} = \frac{1}{\bar{\kappa}} \frac{4\pi \varepsilon_0\varepsilon_r\hbar^2}{\mu e^2} \approx 0.4 \ \mu\text{m} , 
\end{align}
as given in the main text.


\end{document}